\documentclass{aa}  

\usepackage{graphicx}
\usepackage{txfonts}
\usepackage[usenames,dvipsnames]{xcolor}

\newcommand{\veloce}{{\texttt{VELOCE}}}

\begin{document} 

   \title{The VELOCE Modulation Zoo}
   \subtitle{I. Spectroscopic detection of non-radial modes in the first-overtone Cepheids BG~Crucis, QZ~Normae, V0391~Normae, and V0411~Lacertae}


   \author{H. Netzel\thanks{henia@netzel.pl}
          \and
          R. I. Anderson
          \and
          G. Viviani
          }

   \institute{Institute of Physics, \'Ecole Polytechnique F\'ed\'erale de Lausanne (EPFL), Observatoire de Sauverny, 1290 Versoix, Switzerland}

   \date{Received \today; accepted }

 
  \abstract
   {The photometric observations from the recent decade revolutionized our view on classical pulsators. Low-amplitude signals have been detected photometrically in addition to the dominant high-amplitude radial mode pulsations in many RR Lyrae stars and classical Cepheids. First overtone pulsators with an additional low-amplitude signal at a period ratio of around 0.61 with the main mode, the so-called 0.61 stars, form the most populous group among these stars. The nature of this signal has been attributed to non-radial pulsations. Another mysterious group are stars, where the additional signal forms a period ratio of around 0.68 -- the 0.68 stars. The origin of the signal remains unknown.}
   {Here, we search for similar phenomena in spectroscopic observations of first-overtone classical Cepheids collected as part of the \veloce\ project.}
   {We performed frequency analysis of several parameters derived from cross-correlation functions (CCFs), including radial velocity (RV), full width at half maximum (FWHM),  bisector inverse span (BIS), and CCF depth (contrast). Using standard prewhitening, we searched for additional low-amplitude signals. We identify the location of these stars in various sequences of the Petersen diagram.}
   {We detect additional signals in four first-overtone classical Cepheids: BG~Cru, QZ~Nor, V0391~Nor, and V0411~Lac. We classified BG~Cru, QZ~Nor, and V0391~Nor as 0.61 stars based on period ratios. V0411~Lac, however, exhibits a ratio of 0.68 between the two modes, and the additional signal has a longer period. This kind of multiperiodicity remains unexplained.}
   {\veloce\ CCFs yield the first spectroscopic detections of non-radial pulsation modes in classical Cepheids. This opens an asteroseismic window for pursuing a more detailed understanding of these important stars. While the 0.61 signal of BG~Cru, QZ~Nor, V0391~Nor is understood to originate due to non-radial modes of moderate degrees, the 0.68 signal of V0411~Lac still lacks a physical explanation.}


   \maketitle
%

\section{Introduction}

Classical Cepheids are evolved intermediate-mass core-He burning pulsating stars. They pulsate mostly in radial modes with periods from several to a hundred days for the fundamental mode (F) and from around half a day to a few days for the first overtone (1O). Double and triple-mode pulsations in radial modes are also known among Cepheids \citep[see e.g.][]{soszynski2015_blazhko}.

A growing number of additional pulsation phenomena besides pulsations in the dominant radial modes have been reported based on photometric observations in classical Cepheids over the recent decade, including additional low-amplitude signals and period changes. Precise photometric and spectroscopic observations of classical Cepheids also revealed the dominant pulsation period to be unstable on various time scales. A modulation of amplitude and/or phase was detected in classical Cepheids, similarly to the Blazhko effect in RR Lyrae stars \citep{moskalik.kolaczkowski2009,molnar.szabados2014,soszynski2015_blazhko,smolec2017_cep,suveges.anderson2018a,suveges.anderson2018,rathour2021}. 
Space-based photometry from {\it TESS} revealed stochastic cycle-to-cycle variations \citep{plachy2021}. Modulation of pulsation was also shown using interferometry for $\ell$ Car by \cite{anderson2016_interferometry}.

Low-amplitude additional signals are also commonly detected in classical Cepheids in addition to the dominant pulsations in radial modes. \cite{soszynski2008,olech2009,moskalik.kolaczkowski2009} reported in some first-overtone Cepheids an additional signal forming period ratio of around 0.60--0.65 with the first-overtone period. Since the discovery, this group of so-called 0.61 stars significantly grew in members. Most of them were detected in the Magellanic Clouds \citep{soszynski2010,soszynski2015,suveges.anderson2018,smolec2023}. Some 0.61 stars were also found in the Galactic disk \cite{pietrukowicz2013,rathour2021}. Moreover, 0.61 signals are found also in RR Lyrae stars \citep[e.g.][and references therein]{netzel_review}, and in one anomalous Cepheids \citep{plachy2021}. The characteristic feature of 0.61 stars is that they form three nearly parallel sequences in the Petersen diagram (i.e. diagram of shorter-to-longer period ratio vs longer period, see Fig.~\ref{fig:pet}). In many of the 0.61 Cepheids (and RR Lyrae stars) the additional signals are accompanied by their subharmonics, i.e. signals at half of their frequency. \cite{dziembowski2016}, based on calculations of theoretical models for RR Lyrae stars and classical Cepheids, explained the additional signal as harmonics of non-radial modes. That is, the period ratio of around 0.61 is formed by $P_{\rm 1O}/(0.5 P_{\rm nr})$, where $P_{\rm 1O}$ is a first-overtone period and $P_{\rm nr}$ is a non-radial mode period. According to this explanation, sequences in the Petersen diagram, in the case of classical Cepheids, are formed by harmonics of non-radial modes of degrees 7, 8, and 9 for the top, middle, and bottom sequences, respectively \citep[see Petersen diagram in fig. 3 in][]{dziembowski2016}. In the case of RR Lyrae stars, the top and the bottom sequences are formed by harmonics of modes of degrees 8 and 9, respectively \citep[see Petersen diagram in fig. 4 in][]{dziembowski2016}. In other words, subharmonics of the 0.61 signals are in fact non-radial modes according to this explanation. This scenario is currently the most promising. It is supported by the observed properties and distributions of 0.61 signals and their subharmonics for classical Cepheids \cite[e.g.][]{smolec.sniegowska2016} and RR Lyrae stars \citep{netzel_census}. \cite{netzel2022} calculated an extensive grid of theoretical models for RR Lyrae stars with non-radial modes as proposed by \cite{dziembowski2016}, and showed that the non-radial modes are not linearly unstable near the blue edge of the instability strip. Occurrence of the RR Lyrae stars with 0.61 signals in observed color-magnitude diagrams is in agreement with theoretical models. Moreover, the theoretical modeling involving non-radial modes of proposed degrees was carried out and tested using independent observations for RR Lyrae stars and led to promising results \citep{netzel2023}.

However, even though the proposed theory behind the 0.61 stars is supported by the observations and theoretical models, there is still no direct mode identification to verify the theory. Detection of the same signals in spectroscopic observations would be a first step to applying spectroscopic mode identification techniques to unambiguously uncover the nature of signals in the 0.61 stars.

Another well-defined but puzzling multimode group of Cepheids are stars, where the additional signal has a period longer than the first-overtone period, and the corresponding period ratio is around 0.68. First such stars were reported by \cite{suveges.anderson2018}, and more were identified by \cite{smolec2023}. Again, analogous signals are known also in RR Lyrae stars \citep[see a review by][]{netzel_review}. Only recently, the 0.68 signals were identified in fundamental-mode RR Lyrae stars \citep{benko_068}. The nature of the additional signal in 0.68 stars remains unknown, as the considered explanations face difficulties \citep[see discussion in][]{netzel068, benko_068, dziembowski2016}.

Finally, the large-scale study of \cite{suveges.anderson2018} showed the ubiquity of additional signals in classical Cepheids. 


Spectroscopic time series offers new ways to study the modulated signals in classical Cepheids. Analysis of the line profile variations revealed additional variability that cannot be explained by simple pulsation in radial modes in X Sgr, V1334 Cyg, EV Sct and BG Cru \citep{kovtyukh2003}. The presence of non-radial modes was suggested as an explanation. The presence of additional signals on intermediate time scales was reported in Polaris, where line bisector variations exhibit periodic variations with a period of $42$\,d \citep{hatzes.cochran2000}, and possibly of 60\,d \citep{polaris}. 

A long baseline of RV measurements collected as part of the VELOcities of CEpheids project allowed studying modulations of pulsations in Cepheids on different time-scales (Anderson et al. in prep.; henceforth \veloce-I). \citet{anderson2014_qznor} reported that the time scales of modulated signals differ between short-period and long-period stars. Whereas long-timescale variations are seen in short-period, likely first-overtone Cepheids, cycle-to-cycle variations occur in very long-period fundamental mode Cepheids. This dichotomy is also seen among the large sample of \veloce\ Cepheids (\veloce-I). 

Until now, all detections of 0.61 and 0.68 signals were based on photometric observations. \cite{netzel.kolenberg2021} performed a feasibility study of whether it would be possible to detect the non-radial modes of degrees $\ell=7,8,9$ in spectroscopic observations of classical Cepheids and showed that it would be a challenging task requiring a large number of high signal-to-noise (S/N) spectra. However, given the additional signals seen in precision RV data (cf. \veloce-I) and the high signal-to-noise ratio of CCFs, it is likely that Cepheid spectra are equally well suited for detecting these signals.

Here, we analyzed spectroscopic time series data from the \veloce\ project to search for additional periodicities in first-overtone classical Cepheids. This is the first paper of the series on using the spectroscopic time series of \veloce\ to study the signals beyond the dominant pulsation mode and orbital motion of Cepheids -- the modulation zoo. The data used in this study and method of analysis are described in Sec.~\ref{sec:method}. Results are described in Sec.~\ref{sec:results} and discussed in Sec.\ref{sec:discussion}. Sec.~\ref{sec:conclusions} contains conclusions.

\section{Method}\label{sec:method}
The \veloce\ project has collected high-resolution spectroscopic time series observations with baselines exceeding a decade of an unprecedented number of Cepheids from both hemispheres. The details of the data collection for the \veloce\ project can be found in \veloce-I. Here we summarize the most important information about the used dataset. Southern-hemisphere targets are observed with the 1.2m Swiss telescope at La Silla (Chile) equipped with the high-resolution (R$\sim$60\,000) spectrograph Coralie \citep{mayor.queloz1995,segransan2010,vanmalle2016}. The Northern-hemisphere targets are observed with the 1.2m Mercator telescope at Roque de los Muchachos (Spain) observatory equipped with the high-resolution (R$\sim$85\,000) echelle spectrograph Hermes \citep{raskin2011}. The first \veloce\ data release presents over $18,000$ RV observations of 258 Cepheids measured using the cross-correlation technique \citep{Baranne1996}. The resulting CCFs feature extremely high signal-to-noise ratios of around 600-1200 and can be characterized by the parameters RV, FWHM, BIS, and CCF contrast \citep[cf.][for details on these parameters]{anderson2016_lcar}. BIS, in particular, provides a useful measure of CCF asymmetry for time-series analysis. Precise observations together with a long temporal baseline make \veloce\ an excellent database to study rare and hard-to-detect additional phenomena in Cepheids, such as modulations of pulsations or additional low-amplitude periodicities. Observations for \veloce\ are ongoing, and here we used both RV measurements reported in \veloce-I as well as targeted observations acquired beyond the time range included in \veloce-I, i.e. after 5 March 2022 (BJD=2459644).

For Cepheids from the southern hemisphere, we report the results based only on the data collected after November 2014 (\texttt{Coralie14} in \veloce-I, C14 throughout the paper) in order to avoid spurious signals introduced by instrument interventions.
C14 observations provide better line shape stability thanks to the optimal light scrambling properties of the octagonal fibers. Additionally, C14 observations provide up to 9-year temporal baselines with optimal phase sampling for our purposes. 

Hermes spectrograph underwent modifications in April 2018 (cf. \veloce-I). Observations of one of our targets, V0411~Lac, were carried out before 2018 as well. We analyzed the whole dataset and the subset of data collected after modifications separately. The additional signal was detected in both analyses. The results here are presented based on the full dataset.

{\veloce-I\ reports observations for 58\footnote{Classification of 57 stars based on Gaia DR3 classification \citep{ripepi2023}. 58th star is Polaris which is not present in the Gaia classification.} first-overtone Cepheids. 
For our analysis, we used time series for RV and CCF shape indicators: FWHM, BIS, and contrast. RV was determined using a Gaussian profile fit to the CCFs. Line shape indicators were determined from the CCFs themselves. The frequency analysis was performed manually using standard prewhitening. The dominant period was found after applying the Fourier transform. Then, the Fourier series were fitted to the data in the form of:

\begin{equation}
    \label{eq.series}
    F(t)=A_0 + \sum_k A_k \sin(2\pi f_k t + \phi_k),
\end{equation}
where $A_0$ is the mean value, $f_k$ is {\it k}-th frequency, and $A_k$ and $\phi_k$ are its amplitude and phase. We included in the fit the dominant frequency and its harmonics that fulfilled a criterion $A_k/\sigma_k>4.0$. Then, the fit was subtracted from the data, and the Fourier transform was applied to the residuals. Detected signals were included in the fit using Eq.~\ref{eq.series}. The last step was a rejection of outliers using $4\sigma$ clipping. This procedure was repeated for each star for four datasets: RV, FWHM, BIS, and contrast.

Some stars show period changes and/or long-term trends. Period changes manifest in the frequency spectrum as signals unresolved with the first-overtone frequency. Long-term trends manifest as signals in the low-frequency range. We removed long-term trends, regardless of their origin, by subtracting polynomials from the data. QZ Nor shows particularly strong long-term trends, and period changes are present \citep{anderson2014_qznor,anderson2018}. Thus, there are significant remaining signals in the frequency spectrum of the residuals after prewhitening with the dominant frequency and its harmonics. In the case of QZ Nor, we decided to limit the dataset to BJD from 2459331 to 2460157 to limit exposure to long-term trends. QZ Nor is one of the stars with the most observations in \veloce. The selected time range covers the past three years where monitoring has been particularly dense. Hence, by limiting the dataset we avoid the difficulties posed by the long-term modulations while still have sufficiently precise dataset to look for additional low-amplitude signals.

The summary of the data for four stars is presented in Table~\ref{tab:data}.  Table~\ref{tab:data_file} contains the used data for each of the analyzed star.

\begin{table}[]
    \centering
    \begin{tabular}{lllll}
        Star & Instr. & BJD range & <S/N> & N$_{\rm obs}$ \\
        \hline
        BG~Cru & Coralie & 2457154 -- 2460157 & 890 & 405 \\
        V0391~Nor & Coralie & 2457870 -- 2459600 & 250 & 71 \\
        QZ~Nor & Coralie & 2459331 -- 2460157 & 473 & 271 \\
        V0411~Lac & Hermes & 2455742 -- 2460178 & 253 & 91 \\
         & 
    \end{tabular}
    \caption{Summary of the analyzed data for four stars. Consecutive columns provide the star's name, instrument, BJD range, average S/N of CCF profiles calculated using FAMIAS tool \protect\citep{famias}, and the number of observations.}
    \label{tab:data}
\end{table}

\begin{table*}[]
    \centering
    \begin{tabular}{lllllll}
        Star & BJD--2.4M & RV & RV$_{\rm err}$ & FWHM & BIS & Contrast \\
        \hline
BG	Cru	&	57154.63251	&	-19.20946	&	0.009	&	29.85759	&	-0.56679	&	0.60665	\\
BG	Cru	&	57154.72313	&	-18.49692	&	0.009	&	29.97627	&	-0.80856	&	0.6996	\\
BG	Cru	&	57155.64433	&	-13.92248	&	0.01	&	30.0712	&	-1.68633	&	0.20811	\\
BG	Cru	&	57155.72432	&	-14.07354	&	0.01	&	30.09494	&	-1.59799	&	0.09807	\\
BG	Cru	&	57156.75103	&	-23.52224	&	0.009	&	28.90823	&	1.42626	&	-0.92047	\\
BG	Cru	&	57158.6554	&	-14.40592	&	0.01	&	29.71519	&	-1.88471	&	0.63053	\\
BG	Cru	&	57158.75271	&	-14.03372	&	0.014	&	29.69146	&	-1.88858	&	0.37348	\\
BG	Cru	&	57159.74759	&	-20.40442	&	0.011	&	29.83386	&	0.30489	&	-0.80804	\\
BG	Cru	&	57160.61643	&	-23.72161	&	0.009	&	29.38291	&	1.00989	&	-0.51749	\\
\vdots & \vdots & \vdots & \vdots & \vdots & \vdots & \vdots  \\
         & 
    \end{tabular}
    \caption{Sample of a table containing the analyzed data. Consecutive columns provide star's name, BJD of observation, RV with its error, FWHM, BIS, and contrast values. Full table is available online.}
    \label{tab:data_file}
\end{table*}

\section{Results}\label{sec:results}

Phased curves for four used time series and corresponding frequency spectra after prewhitening with the first overtone and its harmonics for the studied stars are presented in Figs.~\ref{fig:bgcru_frequencygrams}, \ref{fig:V0391nor_frequencygrams}, \ref{fig:qznor_frequencygrams}, and \ref{fig:V0411lac_frequencygrams}. In two stars from our sample, the detected additional signals have a period that forms a ratio of 0.60--0.65 with the first-overtone period. These stars are BG Cru and V0391 Nor.

In the case of the BG Cru RV data, after prewhitening, there is still a remaining signal at the position of the first overtone, indicating that the first-overtone amplitude and frequency change in time. The additional signal, forming the period ratio of around 0.61, is detected in the FWHM and contrast data using the arbitrarily chosen threshold of detection of three times the average noise level in the considered frequency range (cyan line in prewhitened power spectra). The signal detected using FWHM data has a S/N=4.24\footnote{S/N reported here for individual peaks is calculated using a box of width $0.5$ c/d and excluding the signal}, so higher than the chosen threshold, whereas the signal detected in the power spectrum of the contrast data has S/N=4.48. Interestingly, there are even more additional signals detected. In the RV, FWHM, and BIS frequency spectra, there is a signal with a period of $P\approx 3.0$ d, which is close to the first-overtone period. Note that this is the only signal remaining in the frequency spectrum of BIS after prewhitening. BG Cru is an interesting object which shows line splitting in their line profiles \citep[see][]{anderson2013, kovtyukh2003}. This observed periodicity might be a manifestation of line splitting and will be studied in detail elsewhere (Netzel et al. in prep). Another additional signal is visible at a low-frequency range for the FWHM and contrast. This signal has a period of around 55\,d. Similar signals in low frequencies with periods of around 42\,d and 60\,d were already reported in Polaris \citep{hatzes.cochran2000,polaris}. The origin of this long-period signal in BG Cru is uncertain.

\begin{figure*}
    \centering
    \includegraphics[width=\textwidth]{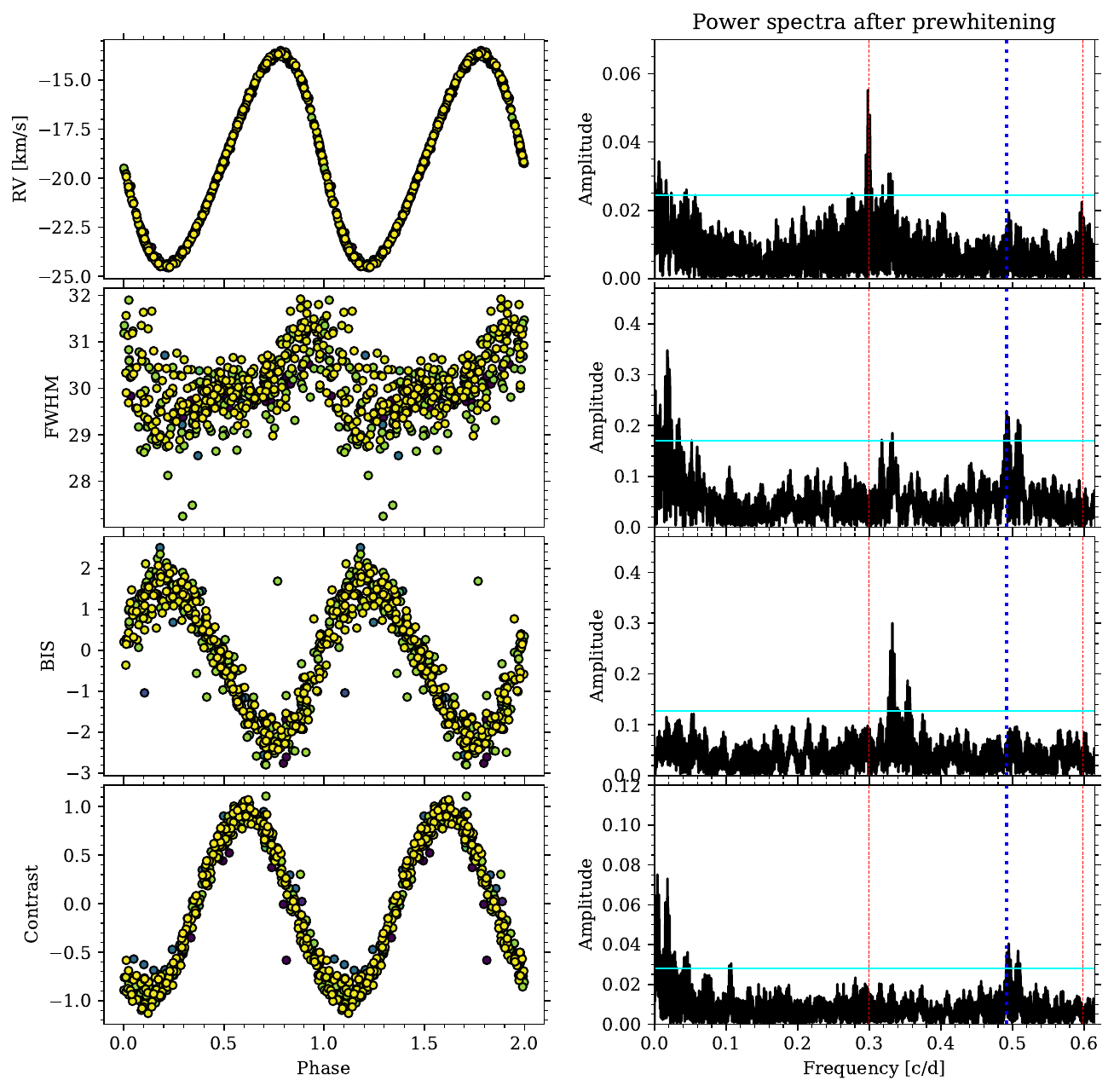}
    \caption{Phased curves and frequency spectra for BG Cru. Consecutive rows correspond to RV, FWHM, BIS and contrast data. Left panels: phased curves. BJD is color-coded. Right panels: frequency spectra after prewhitening with the first overtone and its harmonic. Positions of the first overtone and its harmonic are marked with red dotted lines. Blue dashed line marks the position of the additional signal. Note that its position is marked for datasets where no detection was made. Horizontal cyan line corresponds to three times the average noise level.  The frequency range is plotted up to the Nyquist frequency defined as a median of separations between the observations.}
    \label{fig:bgcru_frequencygrams}
\end{figure*}

\begin{figure*}
    \centering
    \includegraphics[width=\textwidth]{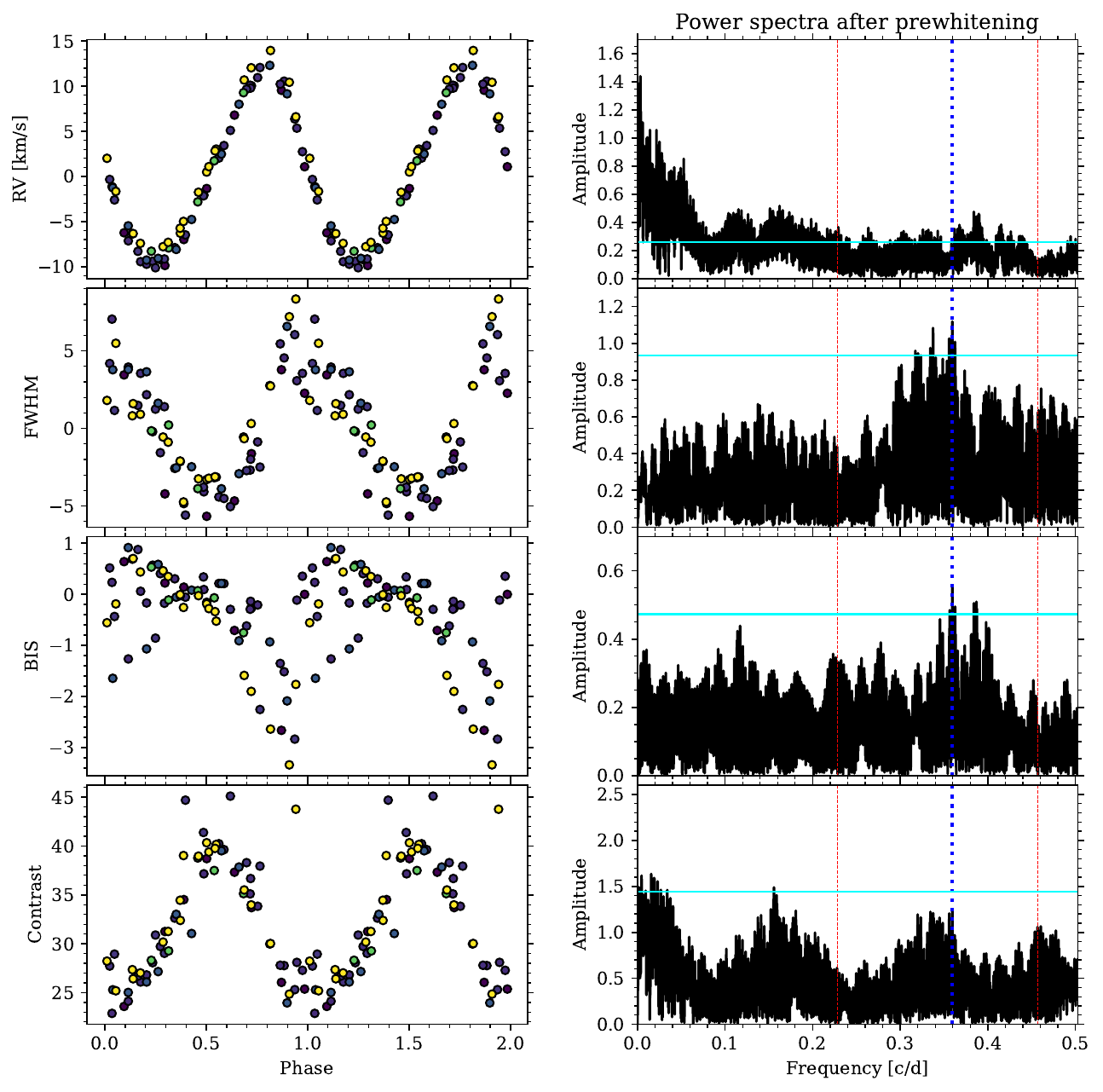}
    \caption{The same as Fig.~\ref{fig:bgcru_frequencygrams} but for V0391 Nor.}
    \label{fig:V0391nor_frequencygrams}
\end{figure*}

In V0391 Nor, the additional signal detected in BIS and FWHM data is not a single coherent peak but rather a cluster of close peaks.  Only few peaks from the cluster exceed the threshold marked with a cyan line. Namely, the S/N of the highest-amplitude signal is S/N=3.51 in the BIS power spectrum. The highest-amplitude signal in the power spectrum of FWHM data has S/N=3.65. In FWHM in particular, the power excess at this frequency range is visible. The period ratio formed by the highest-amplitude peak of the cluster fit nicely the expected period ratio of 0.61 Cepheids. We also do not expect to find any instrumental signals of similar frequency. Hence, we confirm the detection of the signal. We note that similar features in frequency spectra are also known for photometric data for classical Cepheids \citep[see e.g. fig. 7 in][]{rathour2021}, as well as for RR Lyrae stars \citep[e.g.][]{netzel_k2}. It was shown in the case of RR Lyrae stars that the additional signals in 0.61 stars can have strong temporal variations in amplitude and phase, resulting in wide structures observed in frequency spectra \citep[e.g.][]{moskalik2015,benko2023}. Moreover, the observations for V0391 Nor are much less numerous and have significantly lower S/N than for BG Cru (see Table~\ref{tab:data}). In RV data there is a significant trend (visible clearly in the phased RV data) which was not fully removed with a polynomial fitted to the data. Consequently, in the power spectrum there is a strong signal in the low-frequency range. To check whether the increased noise-level might hamper the detection of additional signals, we removed the trend using spline function. After removing trend, no additional signal was detected in the power spectrum of RV.


In QZ Nor, we found an additional signal that has a period longer than the first-overtone period, and they form a period ratio of around 1.28 (shorter-to-longer period ratio of 0.78)  in RV, BIS, and FWHM data. When analyzing the RV data, we detected harmonics of the additional signal and linear combinations between the additional periodicity and the first-overtone frequency. Linear combinations were also found based on BIS data.  In the case of FWHM,  there is an additional signal at the frequency of around 0.47 c/d.  This signal is also detectable in BIS data, but it has a higher amplitude compared to the additional signal forming a period ratio of 0.78. The frequency of 0.47 c/d would correspond to the linear combination of the first-overtone frequency and the frequency of the 0.78 signal. A similar period ratio to that observed in QZ Nor was already reported in the literature as a period ratio formed by the subharmonics of 0.61 signals \citep[based on photometric data, see e.g.][]{rathour2021,smolec2023}. Indeed, the period ratio formed by the harmonic of the additional signal in QZ Nor fits the 0.60--0.65 range. Consequently, we consider QZ Nor also a member of the 0.61 group together with BG Cru and V0391 Nor. We note that among RR Lyrae stars, such cases are also known \citep{netzel_census,benko2021}. QZ Nor also shows long-term RV amplitude changes already studied by \cite{anderson2014_qznor,anderson2020}.  Interestingly, these variations are also visible in FWHM data. Even though we limited the dataset to shorter time range, the changes in FWHM are strong enough to cause the remnant signals unresolved with the first-overtone frequency in the frequency spectrum of FWHM.

\begin{figure*}
    \centering
    \includegraphics[width=\textwidth]{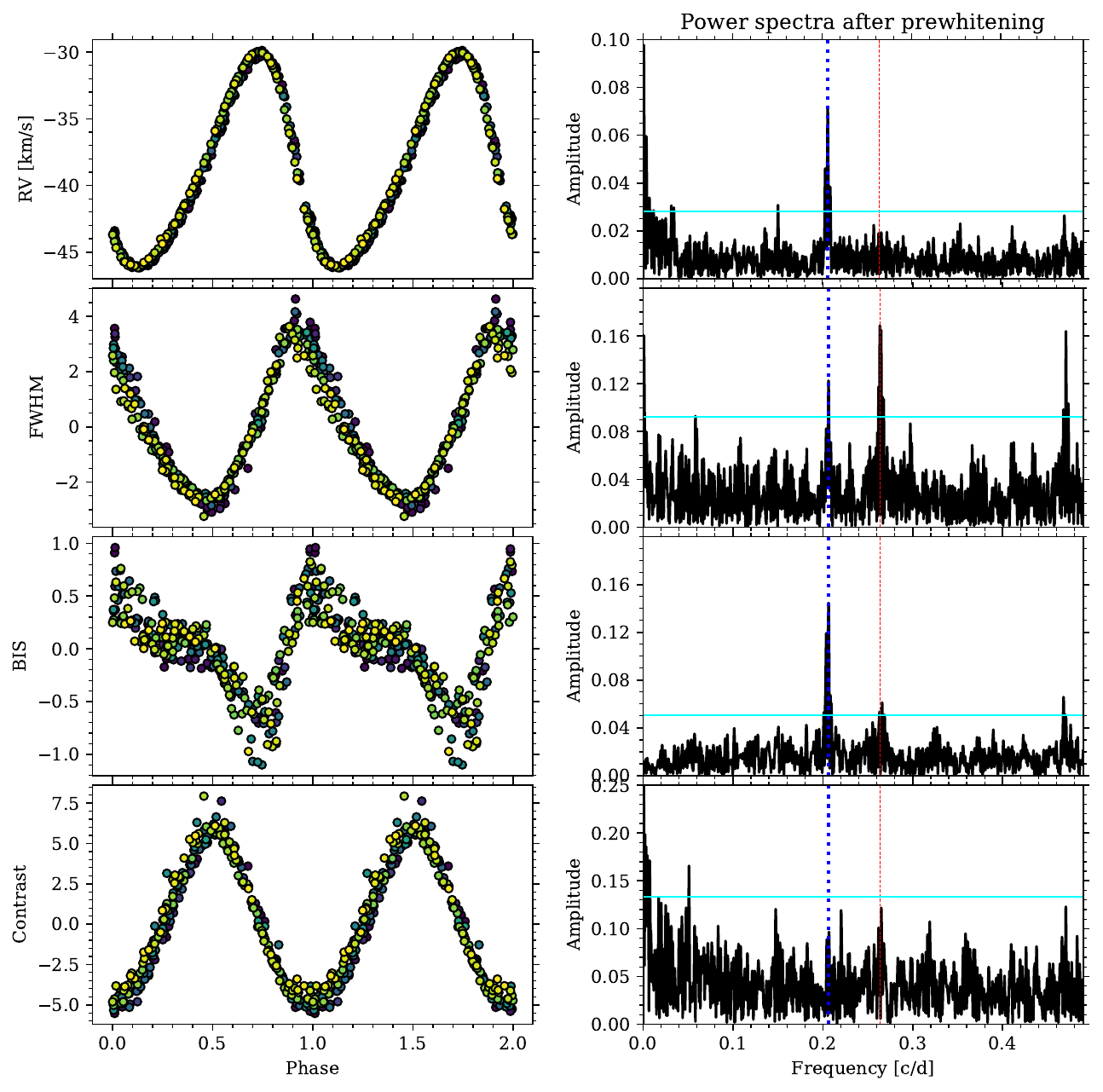}
    \caption{The same as Fig.~\ref{fig:bgcru_frequencygrams} but for QZ Nor. Note that the first-overtone harmonic has higher frequency than the Nyquist frequency and thus is not shown in the power spectrum.}
    \label{fig:qznor_frequencygrams}
\end{figure*}

\begin{figure*}
    \centering
    \includegraphics[width=\textwidth]{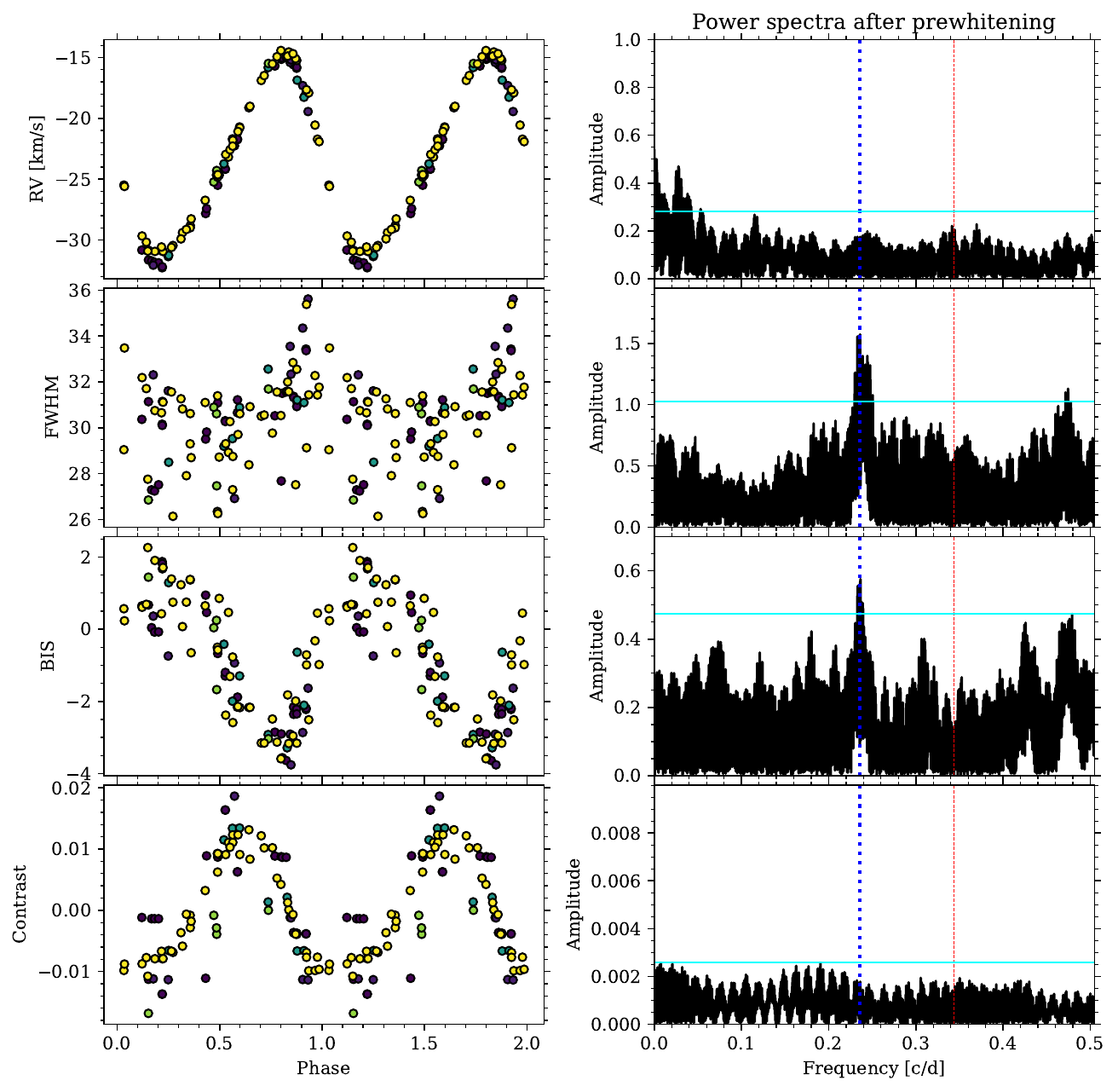}
    \caption{The same as Fig.~\ref{fig:bgcru_frequencygrams} but for V0411 Lac. Note that the first-overtone harmonic has higher frequency than the Nyquist frequency and thus is not shown in the power spectrum.}
    \label{fig:V0411lac_frequencygrams}
\end{figure*}

In V0411 Lac BIS and FWHM data, we detected an additional signal that has a longer period and forms a period ratio of 0.687 with the first-overtone period. The detection is more significant in the FWHM power spectrum, where the additional signals has S/N=4.60. In the BIS power spectrum, the additional signal has S/N=3.68. Moreover, in the FWHM power spectrum, we also detected a harmonic of the additional signal. Based on the period ratio formed by the first-overtone frequency and the additional signal, and given the fact that the additional signal has a longer period than the first overtone, we consider V0411 Lac to be a member of the 0.68 group previously identified using photometric observations.

Periods and period ratios in BG Cru, V0391 Nor, QZ Nor, and V0411 Lac are collected in Table~\ref{tab:results}. Note that there are multiple rows for all stars, because the detections were made in more than one dataset.

\begin{table}[]
    \centering
    \begin{tabular}{l|lll}
    Name & $P_{\rm 1O}$ [d]  & $P_{\rm X}$ [d] & 
    $P_{\rm S}/P_{\rm L}$  \\
    \hline
        BG Cru$_{\rm F}$  &   3.34260(9) &  2.03330(9) &  0.60830 \\
         BG Cru$_{\rm C}$  &   3.34260(1)  &   2.0340(2) &   0.60838 \\
        V0391 Nor$_{\rm F}$  &     4.3727(1)  &   2.7817(2) &   0.63616  \\
        V0391 Nor$_{\rm B}$  &   4.3729(4)   &   2.7817(2) &   0.63612  \\
        QZ Nor$_{\rm R}$  &   3.786803(8) &   4.847(2) &   0.78127 (0.63998)  \\
        QZ Nor$_{\rm F}$  &   3.78659(7) &   4.847(2) &   0.78117 (0.64007)  \\
        QZ Nor$_{\rm B}$  &   3.7872(1)   &   4.846(1) &   0.78158 (0.63973)  \\
        V0411 Lac$_{\rm F}$ &   2.9088(1) &   4.2370(1) &   0.68652 \\
        V0411 Lac$_{\rm B}$ &   2.90847(3) &   4.228(2) &   0.68792 \\
        \
    \end{tabular}
    \caption{Parameters for four stars with detection of additional signals. Consecutive columns provide the name of a star, the first-overtone period, the period of the additional signal, and their shorter-to-longer period ratio. There are multiple rows for stars where detection was made in different datasets: RV (R), FWHM (F), BIS (B) and contrast (C). For QZ Nor, we included in the parenthesis a period ratio that would be formed by the harmonic of the detected additional signal (see text for details).}
    \label{tab:results}
\end{table}

\begin{figure*}
    \centering
    \includegraphics[width=\textwidth]{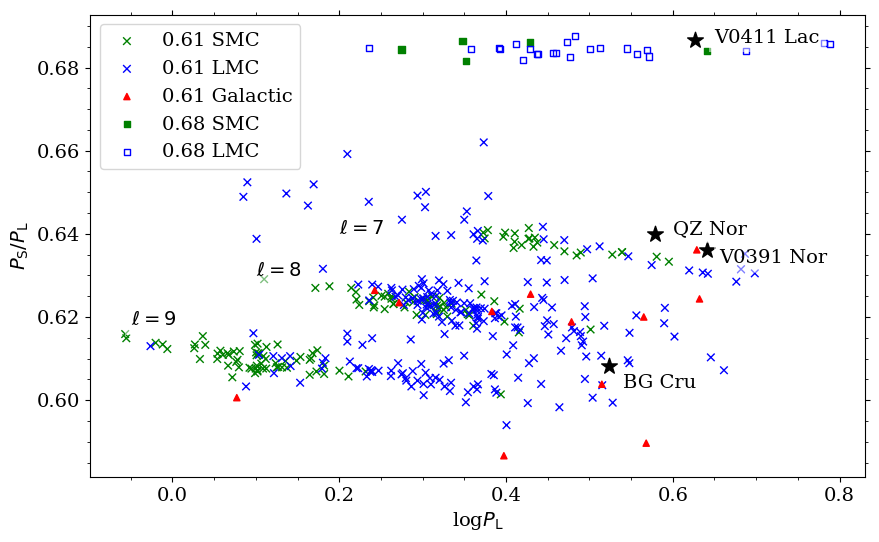}
    \caption{Petersen diagram for 0.61 and 0.68 stars. Green points correspond to Cepheids from SMC; blue points correspond to Cepheids from LMC; red points -- to Cepheids from the Galactic disk \protect\citep{rathour2021}. The stars studied here are plotted with black asterisks using values based on FWHM.  The interpretation of individual sequences, i.e. degree of the non-radial mode, according to the theory by \protect\cite{dziembowski2016} is marked in the figure.}
    \label{fig:pet}
\end{figure*}

We plotted BG Cru, V0391 Nor, QZ Nor, and V0411 Lac in the Petersen diagram (i.e. diagram of period ratio vs the longer period) in Fig.~\ref{fig:pet}. As a reference, we plotted the 0.61 and 0.68 Cepheids identified based on photometric studies. Characteristic three sequences are visible for the 0.61 stars. The 0.68 stars form a single sequence. Note that in the case of 0.61 stars there are differences in sequences between stars from the LMC, SMC and Galactic field. Specifically, the period ratios of individual sequences, dispersion within the sequences and their slope and intercept differ for each sample. Differences between LMC and SMC sample were recently analyzed in detail by \cite{smolec2023} based on photometric observations. The number of known 0.61 stars in the Milky Way (red triangles in Fig.~\ref{fig:pet}) is small. Moreover, they typically have longer periods and show larger scatter.  There is no significant difference between 0.68 stars from LMC and SMC, but the number of known stars is small. 

In the case of 0.61 stars, the theory by \cite{dziembowski2016} predicts that the sequences visible in Fig.~\ref{fig:pet} are formed by the harmonics of non-radial modes of degrees 7, 8 and 9 (see Sec.~\ref{sec:discussion} for further discussion). Degrees, $\ell$, of these modes are indicated in Fig.~\ref{fig:pet} too.

V0391 Nor is located slightly above the long-period part of the top sequence. BG Cru seems to be located in the long-period extension of the middle sequence. However, in the long-period range, the scatter in period ratios is significant, and the sequences are no longer precisely determined as in the case of the short-period range. QZ Nor is plotted in the Petersen diagram using the period ratio that would be formed by the harmonic of the detected additional signal, i.e. $0.5P_{\rm X}/P_{\rm 1O}$, where $P_{\rm X}$ is the period of the detected additional signal.

V0411 Lac fits very well with the progression defined by the other 0.68 stars known from photometric studies.

\section{Discussion}\label{sec:discussion}

Additional short-period signals forming period ratio of around 0.61 are often detected among classical Cepheids and RR Lyrae stars pulsating in the first overtone based on photometric observations. The signals detected in BG Cru and V0391 Nor place these two stars inside the group of 0.61 Cepheids in the Petersen diagram (Fig.~\ref{fig:pet}).

The observed period ratio, that previously was detected only in photometric observations, was considered extensively for Cepheids and RR Lyrae stars. The origin of the additional signal as another radial mode was ruled out because of the value of the period ratio for Cepheids \citep{dziembowski2012}, as well as for RR Lyrae stars \citep{moskalik2015,netzel2015}. Therefore the origin was most likely attributed to non-radial modes. However, the signal forming period ratio of 0.61 was challenging to explain as non-radial modes as discussed by \cite{dziembowski2012}. The most promising explanation, involving subharmonics of the additional signals, was proposed by \cite{dziembowski2016}, and was verified based on photometric observations \citep[e.g.][]{smolec.sniegowska2016}. \cite{netzel.kolenberg2021} discussed the possibility of detecting the same phenomenon as observed in photometric data, but in spectroscopic time-series, and showed that such detection is possible for a dataset of sufficient quality. \cite{netzel.kolenberg2021} also shown, based on synthetic spectra, that the amplitude of the non-radial mode and its harmonic depends on their degree and azimuthal order, as well as on inclination of a star. Hence, in many combinations of these parameters it can be often observed that the harmonic has a higher amplitude than the non-radial mode. This case is similar to what we observe for the 0.61 stars using photometry.

Therefore, in BG Cru and V0391 Nor we most likely observe the same phenomenon. According to the explanation of the 0.61 stars by \cite{dziembowski2016}, the signal detected in BG Cru would correspond to the harmonic of the non-radial mode of degree $\ell=8$. The period ratio in V0391 Nor would correspond to the harmonic of the non-radial mode of degree $\ell=7$.

As already discussed in Sec.~\ref{sec:results}, the signal in QZ Nor is consistent with the interpretation that it is a subharmonic of the 0.61 signal. Therefore, according the explanation by \cite{dziembowski2016} it would correspond to the non-radial mode of degree $\ell=7$. We note, that the case of QZ Nor is not uncommon for 0.61 stars. There are known stars from photometric studies, where the subharmonic has a higher amplitude than the additional signal, or is the only detected additional signal \citep[e.g.][]{smolec2023}. It is worth noting that, in the Petersen diagram, these subharmonics tend to form a clump rather than distinct sequences, in contrast to the well-defined sequences for the 0.61 signals. Moreover, the study of \cite{netzel.kolenberg2021} also showed cases where the non-radial mode has a higher amplitude than its harmonic, or is even the only detected signal.

 The stars studied here fulfill the criteria for the detection established by \cite{netzel.kolenberg2021}. The S/N and the number of spectra required to detect the non-radial modes of degrees 7, 8, and 9 or its harmonics was determined to be at least 50-100 spectra with S/N of at least 200-300. These requirements are fulfilled for the stars analyzed here (see Table~\ref{tab:data}) thanks to the high S/N provided by CCFs. Note however, that in the case of V0391 Nor the dataset falls at the lower limit of the requirements. However, the \veloce\ project is ongoing, and further observations of the analyzed targets are anticipated. This continued data collection should enable a clearer and more definitive identification of the signals discussed in this paper.

\cite{kovtyukh2003} already reported unusual line profile variations in several classical Cepheids, in particular in BG Cru, and suggested pulsations in non-radial modes as a possible explanation. We were able to confirm the presence of additional signals in line profile variations. Moreover, BG Cru shows line splitting \citep{anderson2013,usenko2014}, which will be studied in detail elsewhere (Netzel et al., in prep.).

As visible from Table~\ref{tab:results}, the signals were detected in different datasets. For all stars, the detection was made based on FWHM data. With the exception of BG Cru, the detection was also possible using BIS data. The additional signal was detected using contrast only in the case of BG Cru. QZ Nor is the only star in which the detection was made also based on RV data. It is not clear, what is the reason behind detections and lack of them in different datasets. Note, that used parameters are different measures of line profile variations. RV is sensitive to pulsational velocity, FWHM to turbulent broadening, BIS to both pulsational velocity and velocity gradients, while contrast to temperature variations. Moreover, RV is calculated based on the Gaussian fit to the CCF profile, while FWHM, BIS and contrast are calculated directly from the CCF profile. For all stars these signals were detected in either FWHM or BIS, which are more sensitive to shape variations of CCF profiles. However, in the case of BG Cru, we hypothesize that the line splitting present in this star hampers the detection in BIS.

\veloce\ provides data for 58 first-overtone Cepheids. Here we report the discovery of the additional signals analogous to groups known from photometry in four stars. We note however, that there are significant differences between the datasets for each of the 58 Cepheids in terms of number of observations and signal-to-noise of spectra. Hence, it is not possible to draw any meaningful conclusion on the incidence rate of additional signals in our dataset. We also note, that, the \veloce\ observations continue. Moreover, the sampling of \veloce\ is optimal for detection of such phenomena since around 2021. Hence, we expect to able to increase our sample of first-overtone Cepheids with spectroscopic detections of additional signals. This first result of the series is just a beginning of a systematic study of the modulation zoo of classical Cepheids.

\section{Conclusions}\label{sec:conclusions}

We analyzed spectroscopic time series for first-overtone Cepheids collected during the \veloce\ project (Anderson et al., in prep.). We performed a frequency analysis of radial velocities (RV), full-width at half-maximum (FWHM), bisector inverse span (BIS), and relative depth (contrast) determined from CCFs of very high S/N. Here, we report the very first discovery of signals known in the so-called 0.61 and 0.68 Cepheids in spectroscopic data.

In BG Cru and V0391 Nor, we detected additional signals that form a period ratio with the first overtone in a range of 0.60--0.65, which makes these stars the first members of the 0.61 group detected using spectroscopic data. According to the model by \cite{dziembowski2016}, the signals detected in BG Cru and V0391 Nor correspond to non-radial modes of degrees 7 and 8, respectively. 

In QZ Nor, we detected an additional signal that forms a period ratio of around 1.28 with the first overtone. The harmonic of that signal would form a period ratio of around 0.64. Therefore, we consider QZ Nor to be a member of the 0.61 group, and we note that similar cases are known from photometric studies. The signal detected in QZ Nor is a non-radial mode of degree 7.

In V0411 Lac, we detected an additional signal that has a longer period than the first-overtone period and forms a period ratio of 0.687. We classify this star as a member of the 0.68 group, which was established based on photometric observations. This group is particularly puzzling because the explanation of the nature of the additional signal is yet to be discovered. This very first detection of the 0.68 signal using spectroscopic data will hopefully provide new insights on the origin of this mysterious signal.

In this work, we not only report the discovery of additional periodicity, but we were able to show that, most likely, we deal with the same phenomenon as established based on photometric data for the 0.61 and 0.68 stars. This result takes us one step closer to the concluding identification of the nature of the additional signals.

Our findings show the wealth of information that can be derived from line shape indicators such as full width at half maximum, bisector inverse span and line relative depth, in addition to the insights offered by radial velocity time series.

\begin{acknowledgements}
This work was supported by the European Research Council (ERC) under the European Union’s Horizon 2020 research and innovation programme (Grant Agreement No. 947660). RIA is funded by the SNSF through an Eccellenza Professorial Fellowship, grant number PCEFP2\_194638.

This work uses frequency analysis software written by R. Smolec.

The Euler telescope is funded by the Swiss National Science Foundation (SNSF).

This research is based on observations made with the Mercator Telescope, operated on the island of La Palma by the Flemish Community, at the Spanish Observatorio del Roque de los Muchachos of the Instituto de Astrof\'isica de Canarias. {\it Hermes} is supported by the Fund for Scientific Research of Flanders (FWO), Belgium, the Research Council of K.U. Leuven, Belgium, the Fonds National de la Recherche Scientifique (F.R.S.-FNRS), Belgium, the Royal Observatory of Belgium, the Observatoire de Gene\`eve, Switzerland, and the Th\"uringer Landessternwarte, Tautenburg, Germany.

We acknowledge the contributions of all observers who contributed to collecting the VELOCE dataset.
\end{acknowledgements}

\bibliographystyle{aa}
\bibliography{biblio.bib}

\begin{thebibliography}{46}
\expandafter\ifx\csname natexlab\endcsname\relax\def\natexlab#1{#1}\fi

\bibitem[{{Anderson}(2013)}]{anderson2013}
{Anderson}, R.~I. 2013, PhD thesis, University of Geneva, Astronomical
  Observatory

\bibitem[{{Anderson}(2014)}]{anderson2014_qznor}
{Anderson}, R.~I. 2014, \aap, 566, L10

\bibitem[{{Anderson}(2016)}]{anderson2016_lcar}
{Anderson}, R.~I. 2016, \mnras, 463, 1707

\bibitem[{{Anderson}(2018)}]{anderson2018}
{Anderson}, R.~I. 2018, in The RR Lyrae 2017 Conference. Revival of the
  Classical Pulsators: from Galactic Structure to Stellar Interior Diagnostics,
  ed. R.~{Smolec}, K.~{Kinemuchi}, \& R.~I. {Anderson}, Vol.~6, 193--200

\bibitem[{{Anderson}(2019)}]{polaris}
{Anderson}, R.~I. 2019, \aap, 623, A146

\bibitem[{{Anderson}(2020)}]{anderson2020}
{Anderson}, R.~I. 2020, in Stars and their Variability Observed from Space, ed.
  C.~{Neiner}, W.~W. {Weiss}, D.~{Baade}, R.~E. {Griffin}, C.~C. {Lovekin}, \&
  A.~F.~J. {Moffat}, 61--66

\bibitem[{{Anderson} {et~al.}(2016){Anderson}, {M{\'e}rand}, {Kervella},
  {Breitfelder}, {LeBouquin}, {Eyer}, {Gallenne}, {Palaversa}, {Semaan},
  {Saesen}, \& {Mowlavi}}]{anderson2016_interferometry}
{Anderson}, R.~I., {M{\'e}rand}, A., {Kervella}, P., {et~al.} 2016, \mnras,
  455, 4231

\bibitem[{{Baranne} {et~al.}(1996){Baranne}, {Queloz}, {Mayor}, {Adrianzyk},
  {Knispel}, {Kohler}, {Lacroix}, {Meunier}, {Rimbaud}, \& {Vin}}]{Baranne1996}
{Baranne}, A., {Queloz}, D., {Mayor}, M., {et~al.} 1996, \aaps, 119, 373

\bibitem[{{Benk{\H{o}}} \& {Kov{\'a}cs}(2023)}]{benko_068}
{Benk{\H{o}}}, J.~M. \& {Kov{\'a}cs}, G.~B. 2023, arXiv e-prints,
  arXiv:2311.11670

\bibitem[{{Benk{\H{o}}} {et~al.}(2023){Benk{\H{o}}}, {Plachy}, {Netzel},
  {B{\'o}di}, {Moln{\'a}r}, \& {P{\'a}l}}]{benko2023}
{Benk{\H{o}}}, J.~M., {Plachy}, E., {Netzel}, H., {et~al.} 2023, \mnras, 521,
  443

\bibitem[{{Benk{\H{o}}} {et~al.}(2021){Benk{\H{o}}}, {S{\'o}dor}, \&
  {P{\'a}l}}]{benko2021}
{Benk{\H{o}}}, J.~M., {S{\'o}dor}, {\'A}., \& {P{\'a}l}, A. 2021, \mnras, 500,
  2554

\bibitem[{{Dziembowski}(2012)}]{dziembowski2012}
{Dziembowski}, W.~A. 2012, \actaa, 62, 323

\bibitem[{{Dziembowski}(2016)}]{dziembowski2016}
{Dziembowski}, W.~A. 2016, Commmunications of the Konkoly Observatory Hungary,
  105, 23

\bibitem[{{Hatzes} \& {Cochran}(2000)}]{hatzes.cochran2000}
{Hatzes}, A.~P. \& {Cochran}, W.~D. 2000, \aj, 120, 979

\bibitem[{{Kovtyukh} {et~al.}(2003){Kovtyukh}, {Andrievsky}, {Luck}, \&
  {Gorlova}}]{kovtyukh2003}
{Kovtyukh}, V.~V., {Andrievsky}, S.~M., {Luck}, R.~E., \& {Gorlova}, N.~I.
  2003, \aap, 401, 661

\bibitem[{{Mayor} \& {Queloz}(1995)}]{mayor.queloz1995}
{Mayor}, M. \& {Queloz}, D. 1995, \nat, 378, 355

\bibitem[{{Moln{\'a}r} \& {Szabados}(2014)}]{molnar.szabados2014}
{Moln{\'a}r}, L. \& {Szabados}, L. 2014, \mnras, 442, 3222

\bibitem[{{Moskalik} \& {Ko{\l}aczkowski}(2009)}]{moskalik.kolaczkowski2009}
{Moskalik}, P. \& {Ko{\l}aczkowski}, Z. 2009, \mnras, 394, 1649

\bibitem[{{Moskalik} {et~al.}(2015){Moskalik}, {Smolec}, {Kolenberg},
  {Moln{\'a}r}, {Kurtz}, {Szab{\'o}}, {Benk{\H{o}}}, {Nemec}, {Chadid},
  {Guggenberger}, {Ngeow}, {Jeon}, {Kopacki}, \& {Kanbur}}]{moskalik2015}
{Moskalik}, P., {Smolec}, R., {Kolenberg}, K., {et~al.} 2015, \mnras, 447, 2348

\bibitem[{Netzel(2023)}]{netzel_review}
Netzel, H. 2023, Blazhko effect and the Petersen diagram

\bibitem[{{Netzel} \& {Kolenberg}(2021)}]{netzel.kolenberg2021}
{Netzel}, H. \& {Kolenberg}, K. 2021, \mnras, 508, 3508

\bibitem[{{Netzel} {et~al.}(2023{\natexlab{a}}){Netzel}, {Moln{\'a}r}, \&
  {Joyce}}]{netzel2023}
{Netzel}, H., {Moln{\'a}r}, L., \& {Joyce}, M. 2023{\natexlab{a}}, \mnras, 525,
  5378

\bibitem[{{Netzel} {et~al.}(2023{\natexlab{b}}){Netzel}, {Moln{\'a}r},
  {Plachy}, \& {Benk{\H{o}}}}]{netzel_k2}
{Netzel}, H., {Moln{\'a}r}, L., {Plachy}, E., \& {Benk{\H{o}}}, J.~M.
  2023{\natexlab{b}}, \aap, 677, A177

\bibitem[{{Netzel} \& {Smolec}(2019)}]{netzel_census}
{Netzel}, H. \& {Smolec}, R. 2019, \mnras, 487, 5584

\bibitem[{{Netzel} \& {Smolec}(2022)}]{netzel2022}
{Netzel}, H. \& {Smolec}, R. 2022, \mnras, 515, 3439

\bibitem[{{Netzel} {et~al.}(2015{\natexlab{a}}){Netzel}, {Smolec}, \&
  {Dziembowski}}]{netzel068}
{Netzel}, H., {Smolec}, R., \& {Dziembowski}, W. 2015{\natexlab{a}}, \mnras,
  451, L25

\bibitem[{{Netzel} {et~al.}(2015{\natexlab{b}}){Netzel}, {Smolec}, \&
  {Moskalik}}]{netzel2015}
{Netzel}, H., {Smolec}, R., \& {Moskalik}, P. 2015{\natexlab{b}}, \mnras, 447,
  1173

\bibitem[{{Olech} \& {Moskalik}(2009)}]{olech2009}
{Olech}, A. \& {Moskalik}, P. 2009, \aap, 494, L17

\bibitem[{{Pietrukowicz} {et~al.}(2013){Pietrukowicz}, {Dziembowski},
  {Mr{\'o}z}, {Soszy{\'n}ski}, {Udalski}, {Poleski}, {Szyma{\'n}ski}, {Kubiak},
  {Pietrzy{\'n}ski}, {Wyrzykowski}, {Ulaczyk}, {Koz{\l}owski}, \&
  {Skowron}}]{pietrukowicz2013}
{Pietrukowicz}, P., {Dziembowski}, W.~A., {Mr{\'o}z}, P., {et~al.} 2013,
  \actaa, 63, 379

\bibitem[{{Plachy} {et~al.}(2021){Plachy}, {P{\'a}l}, {B{\'o}di}, {Szab{\'o}},
  {Moln{\'a}r}, {Szabados}, {Benk{\H{o}}}, {Anderson}, {Bellinger}, {Bhardwaj},
  {Ebadi}, {Gazeas}, {Hambsch}, {Hasanzadeh}, {Jurkovic}, {Kalaee}, {Kervella},
  {Kolenberg}, {Miko{\l}ajczyk}, {Nardetto}, {Nemec}, {Netzel}, {Ngeow},
  {Ozuyar}, {Pascual-Granado}, {Pilecki}, {Ripepi}, {Skarka}, {Smolec},
  {S{\'o}dor}, {Szab{\'o}}, {Christensen-Dalsgaard}, {Jenkins}, {Kjeldsen},
  {Ricker}, \& {Vanderspek}}]{plachy2021}
{Plachy}, E., {P{\'a}l}, A., {B{\'o}di}, A., {et~al.} 2021, \apjs, 253, 11

\bibitem[{{Raskin} {et~al.}(2011){Raskin}, {van Winckel}, {Hensberge},
  {Jorissen}, {Lehmann}, {Waelkens}, {Avila}, {de Cuyper}, {Degroote},
  {Dubosson}, {Dumortier}, {Fr{\'e}mat}, {Laux}, {Michaud}, {Morren}, {Perez
  Padilla}, {Pessemier}, {Prins}, {Smolders}, {van Eck}, \&
  {Winkler}}]{raskin2011}
{Raskin}, G., {van Winckel}, H., {Hensberge}, H., {et~al.} 2011, \aap, 526, A69

\bibitem[{{Rathour} {et~al.}(2021){Rathour}, {Smolec}, \&
  {Netzel}}]{rathour2021}
{Rathour}, R.~S., {Smolec}, R., \& {Netzel}, H. 2021, \mnras, 505, 5412

\bibitem[{{Ripepi} {et~al.}(2023){Ripepi}, {Clementini}, {Molinaro}, {Leccia},
  {Plachy}, {Moln{\'a}r}, {Rimoldini}, {Musella}, {Marconi}, {Garofalo},
  {Audard}, {Holl}, {Evans}, {Jevardat de Fombelle}, {Lecoeur-Taibi},
  {Marchal}, {Mowlavi}, {Muraveva}, {Nienartowicz}, {Sartoretti}, {Szabados},
  \& {Eyer}}]{ripepi2023}
{Ripepi}, V., {Clementini}, G., {Molinaro}, R., {et~al.} 2023, \aap, 674, A17

\bibitem[{{S{\'e}gransan} {et~al.}(2010){S{\'e}gransan}, {Udry}, {Mayor},
  {Naef}, {Pepe}, {Queloz}, {Santos}, {Demory}, {Figueira}, {Gillon},
  {Marmier}, {M{\'e}gevand}, {Sosnowska}, {Tamuz}, \& {Triaud}}]{segransan2010}
{S{\'e}gransan}, D., {Udry}, S., {Mayor}, M., {et~al.} 2010, \aap, 511, A45

\bibitem[{{Smolec}(2017)}]{smolec2017_cep}
{Smolec}, R. 2017, \mnras, 468, 4299

\bibitem[{{Smolec} \& {{\'S}niegowska}(2016)}]{smolec.sniegowska2016}
{Smolec}, R. \& {{\'S}niegowska}, M. 2016, \mnras, 458, 3561

\bibitem[{{Smolec} {et~al.}(2023){Smolec}, {Zi{\'o}{\l}kowska}, {Ochalik}, \&
  {{\'S}niegowska}}]{smolec2023}
{Smolec}, R., {Zi{\'o}{\l}kowska}, O., {Ochalik}, M., \& {{\'S}niegowska}, M.
  2023, \mnras, 519, 4010

\bibitem[{{Soszy{\'n}ski} {et~al.}(2008){Soszy{\'n}ski}, {Poleski}, {Udalski},
  {Szymanski}, {Kubiak}, {Pietrzynski}, {Wyrzykowski}, {Szewczyk}, \&
  {Ulaczyk}}]{soszynski2008}
{Soszy{\'n}ski}, I., {Poleski}, R., {Udalski}, A., {et~al.} 2008, \actaa, 58,
  163

\bibitem[{{Soszy{\'n}ski} {et~al.}(2010){Soszy{\'n}ski}, {Poleski}, {Udalski},
  {Szyma{\'n}ski}, {Kubiak}, {Pietrzy{\'n}ski}, {Wyrzykowski}, {Szewczyk}, \&
  {Ulaczyk}}]{soszynski2010}
{Soszy{\'n}ski}, I., {Poleski}, R., {Udalski}, A., {et~al.} 2010, \actaa, 60,
  17

\bibitem[{{Soszy{\'n}ski} {et~al.}(2015{\natexlab{a}}){Soszy{\'n}ski},
  {Udalski}, {Szyma{\'n}ski}, {Poleski}, {Pietrukowicz}, {Koz{\l}owski},
  {Mr{\'o}z}, {Wyrzykowski}, {Skowron}, {Skowron}, {Pietrzy{\'n}ski},
  {Ulaczyk}, \& {Pawlak}}]{soszynski2015_blazhko}
{Soszy{\'n}ski}, I., {Udalski}, A., {Szyma{\'n}ski}, M.~K., {et~al.}
  2015{\natexlab{a}}, \actaa, 65, 329

\bibitem[{{Soszy{\'n}ski} {et~al.}(2015{\natexlab{b}}){Soszy{\'n}ski},
  {Udalski}, {Szyma{\'n}ski}, {Skowron}, {Pietrzy{\'n}ski}, {Poleski},
  {Pietrukowicz}, {Skowron}, {Mr{\'o}z}, {Koz{\l}owski}, {Wyrzykowski},
  {Ulaczyk}, \& {Pawlak}}]{soszynski2015}
{Soszy{\'n}ski}, I., {Udalski}, A., {Szyma{\'n}ski}, M.~K., {et~al.}
  2015{\natexlab{b}}, \actaa, 65, 297

\bibitem[{{S{\"u}veges} \&
  {Anderson}(2018{\natexlab{a}})}]{suveges.anderson2018}
{S{\"u}veges}, M. \& {Anderson}, R.~I. 2018{\natexlab{a}}, \mnras, 478, 1425

\bibitem[{{S{\"u}veges} \&
  {Anderson}(2018{\natexlab{b}})}]{suveges.anderson2018a}
{S{\"u}veges}, M. \& {Anderson}, R.~I. 2018{\natexlab{b}}, \aap, 610, A86

\bibitem[{{Usenko} {et~al.}(2014){Usenko}, {Kniazev}, {Berdnikov}, {Fokin}, \&
  {Kravtsov}}]{usenko2014}
{Usenko}, I.~A., {Kniazev}, A.~Y., {Berdnikov}, L.~N., {Fokin}, A.~B., \&
  {Kravtsov}, V.~V. 2014, Astronomy Letters, 40, 435

\bibitem[{{Van Malle}(2016)}]{vanmalle2016}
{Van Malle}, M.~N. 2016, PhD thesis, University of Geneva, Switzerland

\bibitem[{{Zima}(2008)}]{famias}
{Zima}, W. 2008, Communications in Asteroseismology, 157, 387

\end{thebibliography}

\end{document}